\documentstyle[11pt]{article}
\normalbaselines

\def\R{I\!\! R}
\def\t{\tau}
\def\ft{Fourier transform\ }
\def\Qt{\widetilde{Q}}
\def\Tr{\mbox{Tr}}
\def\Pib{{\bf \Pi}}

\def\pd#1#2{\frac{\partial #1}{\partial #2} }

\def\l{\sigma}
\def\Ab{A}
\def\Db{{\bf D}}
\def\D{{\bf D}}
\def\Pb{{\bf P }}
\def\Yb{{\bf Y }}
\def\W{{\bf W }}

\def\Qb{{\bf Q }}

\def\ket#1{|#1 \rangle}
\def\bra#1{\langle #1|}
\def\sp #1{\langle #1 \rangle}

\newcommand{\intt}{\int \!\!\!\int}
\newcommand{\be}{\begin{equation}}
\newcommand{\ee}{\end{equation}}
\newcommand{\el}[1]{\label{#1} \end{equation}}
\newcommand{\erl}[1]{\label{#1} \end{eqnarray}}
\newcommand{\lb}[1]{\label{#1}}
\newcommand{\rf}[1]{(\ref{#1})}
\newcommand{\br}{\begin{eqnarray}}
\newcommand{\er}{\end{eqnarray}}

\newcommand{\nn}{\nonumber}
\newcommand{\for}{\qquad \mbox{ for}\quad}

\newcommand{\where}{\quad \mbox{ where}\quad}

\newcommand{\hrt}{\heartsuit}

\begin{document}
\bibliographystyle{unsrt}

\begin{flushright} 
hep-th/9601127 \\ January 1996
\end{flushright}

\begin{center}
{\LARGE \bf Forward-Backward Squeezing Propagator}\\[5mm]

 Jamil \ Daboul \footnote {E-mail: daboul@bguvms.bgu.ac.il} \\[3mm]
{\it Physics Department, Ben Gurion University of the Negev\\
84105 Beer Sheva, Israel} \\
\end{center}

\begin{abstract}
I show that a usual propagator  cannot be defined for the pseudo-diffusion
equation of the Q-functions. Instead, 
a forward-backward propagator is defined, which motivated
a generalization of Cahill-Glauber interpolating operator. 
An algorithm is also given for squeezing Q 
functions directly, using one-dimensional diffusion propagators.
 
\end{abstract}

\section{Introduction}

In previous papers \cite{md,dab}, it was shown
that the Q functions for an operator $A$ (see definition in Eq. \rf{qf} 
below) obeys the following partial differential equation
\be 
\heartsuit(p,q;\l)\; Q(A:p,q;\l) :=  \left [ \pd{}{\l}
 - \frac{1}{4} \left( \frac{\partial^{2}}{\partial q^{2}}
- \frac{1}{\l^{2}} \frac{\partial^{2}}{\partial p^2} \right)\right]
Q(A:p,q;\l)=0~, \where \l:=e^{2y}~,  \lb{psd}
\ee
 where $y$ is the squeezing parameter, as defined in
\rf{A6}. (Note that $\l$ here is the inverse of the $\lambda:=e^{-2y}$ in 
\cite{md,dab}, so that the roles of $p$ and $q$ get exchanged.)
This equation describes how the Q functions $Q(p,q;\l)$ get changed 
in phase space $(p,q)$ as the squeezing parameter $\l$ is increased. 
If the Q function  belongs to a density operator $\rho$, then 
$Q(\rho:p,q;\l)$ is a probability distribution function, which
remains normalized to 1 for all $\l$, i.e. 
if its integral is carried out over the whole phase space. 
Eq. \rf{psd} was called {\em pseudo-diffusion equation} \cite{md,dab}, 
because (a) it resembles the 
diffusion equation in 2 dimensions \cite{widder}, where the parameter $\l$
plays the role of time, and (b) the
coefficients of $\frac{\partial^{2}}{\partial p^{2}}$ and 
$\frac{\partial^{2}
}{\partial q^2}$ in (\ref{psd}) have opposite signs. (Since $\l$ is a 
monotonically {\em increasing} function of $y$, I shall use the time 
analogy 
when referring to either $\l$ or $y$, and hope that this will not lead 
to any confusion). Therefore, this
equation describes a diffusive process in the $q$ variable and an infusive
one in the $p$ variable {\em for all $\l $}. In this way a thin
distribution along the $p$-axis get continuously deformed into a thin
distribution along the $q$-axis, as $\l $ is increased from $0$ to 
$\infty$. 

The (symplectic) Fourier transform of the above partial differential 
equation \rf{psd},
with respect to the phase-space variables $(p,q)$,  yields
an ordinary differential equation in $\l$:
\be
\left[ \frac{\partial }{\partial \l} +\frac{1}{4}\left(k^2-\frac{1}
{\l^2}x^2 \right) \right] \Qt(k,x;\l)=0~,
\el{psdf}
where  the symplectic \ft is defined by
\be \Qt(k,x;\l) := \intt   \frac {dp dq}
{2 \pi} e^{-i [x p- k q]}\; Q(p,q; \l)~.
 \el{sft}
It is easy to check that the {\em `` kernel "}
\be
K(k,x;\l,\mu) :=
 e^{- \frac{1}{4}k^2 (\l-\mu)} e^{- \frac{1}{4} x^2
(\l^{-1}-\mu^{-1})}~,  \el{kern}
is a solution of \rf{psdf}.
Hence, the inverse \ft of the products $K\Qt$
yields the solution of the pseudo-diffusion equation \rf{psd}, for any 
given initial Q function $Q(p,q;\mu)$ with $\mu \le \l$:
\br
Q(p,q;\l) &=& \intt \frac{dk dx}{2 \pi} e^{i [x p- k q]} 
K(k,x;\l,\mu)\; \Qt(k,x; \mu) .  \nn \\
 &=& \intt \frac{dk dx}{2 \pi} e^{i [x p- k q]} 
K(k,x;\l,\mu) 
\intt \frac{dp' dq'}{2 \pi} e^{-i [x p'- k q']} 
\; Q(p',q'; \mu)~ . 
\erl{fs}
Eq. \rf{fs} shows that in order to get the solutions $Q(p,q;\l)$
 we must perform {\em two} double integrations, first over $(p',q')$ and
then over $(k,x)$. Therefore, it would have been much more efficient,
if we are allowed to exchange the order of two double
integrations, and carry out the integration over the variables $(k,x)$ 
first: the resulting integral, if it existed, would have been equal to
 the \ft of the kernel $K$: 
\br
 G_0(p-p',q-q';\l,\mu) &:=& \intt
 \frac{dk dx}{2 \pi}\; e^{i [x (p-p')-k (q-q')]}
\; K(k,x;\l,\mu) \nn \\
&=& 
\intt  \frac{dk dx}{2 \pi}\;  e^{i [x (p-p')- 
k (q-q')]} \; e^{- \frac{1}{4}[k^2 (\l-\mu)+ x^2 (\l^{-1}-\mu^{-1})]}~,
\erl{ftk}
and would have enabled us to calculate the squeezed
distributions (from unsqueezed or less squeezed ones) directly 
by a {\em single} double integration, as follows
\be 
Q(p,q;\l)= \intt   dp' dq' \; G_0(p-p',q-q';\l,\mu)\; Q(p',q';\mu)~,
 \for \l>\mu~.
\ee
Unfortunately, the integral \rf{ftk} does not exist, as we 
shall see in section 3. Nevertheless, I shall give an algorithm for
calculating squeezed Q functions which is based on seperation of variables
and uses 1-dimensional diffusion propagators.

Next, I show that it is more natural to define 2-sided propagators 
\rf{prop} for the pseudo-diffusion equation \rf{psd}; these 
propagators depend on three squeezing parameters,
{\em the present $\l$, the past $\mu \le \l$ and the future $\lambda \ge 
\l$}.
Therefore, I shall refer to these 2-sided propagators as
{\em forward-backward (or future-past) 
squeezing propagators}. A special case of these propagators connects
Q functions to their Wigner counterparts. This makes it possible
to give a new interpretation of the Wigner function. In turn, this 
interpretation makes it sensible to define formally
{\em generalized Q functions and \Qb operators}
which depend on two squeezing parameters, $\l_p$ and $\l_q$. On a subdomain
of these parameters, which can be expressed
in terms of an ordinary squeezing parameter $\l$ and a 
``thermal parameter" $\t$, the above generalized  \Qb operators can be 
identified with thermalized and squeezed Wigner operators  (I hope to
give details elsewhere). Partial 
differentiation of these operators with respect to 
$\l$ and $\t$ yields the generalized pseudo-diffusion equation \rf{gpsd}
and the diffusion equation \rf{tpde}, respectively.
 
In this paper, I review in Sec.~2 basic properties of the Wigner and 
the Q functions.
Then I explain in Sec.~3 why the \ft of the kernel $K$ cannot exist.
In Sec.~4 I give an algorithm for calculating squeezed Q functions.
In Sec.~5 I define the two-sided
propagators for the pseudo-diffusion equation. In Sec.~6 
I give the new interpretation of the Wigner function.
In Sec.~7 I define and study generalized Q functions and \Qb operators.
Finally, in Sec.~8 I give a summary.

\section{Wigner and  Q functions}

The Wigner representation of an operator $\Ab$ is defined formally by 
\cite{kim}
\be
W (A: p, q) := 2\int_{-\infty }^\infty  (q-x\mid \Ab\mid q+x) \ e^{2ixp}dx 
= 2\ \Tr \,(\, \Ab\,  \W(p,q)\, )~,  \lb{wig}
\ee
where I use the round kets to denote eigenstates
of the position operator, ${\bf Q} \mid \! x)= x \mid \! x)$, to 
distinguish them from the number states $\ket{n}$. 
The operators (one for each $(p,q)$)
\be \W(p,q):= 
\int_{-\infty }^\infty  \mid \! q+x)(q-x\!\mid \;  e^{2ixp} \ dx ~,
 \el{wo} 
were called {\em Wigner operators} \ by 
Dahl \cite{dahl}.  Obviously,
\be \W(0,0)= \int_{-\infty }^\infty  \mid \! x)(-x\!\mid  \ dx ~,
 \ee 
is the parity operator. Therefore, the Wigner operators  
can be interpreted as {\em displaced parity operators} [5-8], because
\be \W(p,q) = \Db(p,q) \W(0,0) \Db^\dagger(p,q) =
 \Db(2p,2q) \W(0,0) ~,\el{wd}
where 
\be
\Db(p,q) := 
\exp\ [i (p{\bf Q}-q{\bf P})]= 
\exp\  [\frac i{2}pq] \exp\ [-i q \Pb] \exp\ [ip \Qb]
\el{d}
is called the {\em displacement operator}, because
 $\Db(p,q)\mid \! x)= e^{ip(x+q/2)}\mid \! x+q)  $.
Since $\W(0,0)$ is unitary and Hermitian, it follows immediately 
from \rf{wd} that also the displaced parity operators are unitary and
Hermitian: $ \W^{-1}(p,q)= \W^\dagger(p,q)= \W(p,q)$, so that
$ \W^2(p,q)= {\bf I}$. This means that the $ \W(p,q)$ are observables with 
eigenvalues $\pm 1$, as was pointed out by Dahl \cite{dahl} and
Bishop and Vourdas \cite{bv}.

The Wigner representation \rf{wig} yields functions of two 
variables, $p$ and $q$, 
which may be looked upon as phase-space variables. These 
{\em `Wigner functions'} have  interesting properties and are useful for 
various calculations \cite{kim}. They are often referred to as 
quasi-probability functions,
because they can take negative values  even when $\Ab$ is a positive 
operator, $\Ab\ge 0$, such as a density operator $\rho$.

In contrast, the Q representation
yield nonnegative functions for positive operators $\Ab$: These functions
are defined as follows  \cite{kim,nieto} 
\be
Q(\Ab:p,q;\zeta)=\langle pq;\zeta\mid \Ab \mid pq;\zeta\rangle =
 \Tr\,(\,\Ab\ \Pib(pq;\zeta)\,)~,
\where \Pib(p,q;\zeta) := \mid
pq;\zeta\rangle \langle pq;\zeta\! \mid  \lb{qf}
\ee
are projection operators on the squeezed states $\ket{pq;\zeta}$, which
are defined by \cite{kim,nieto}
\be
|pq;\zeta \rangle ={\bf D}(p,q) {\bf S}(\zeta)|0 \rangle~, \quad 
\mbox{where}\quad \zeta:= ye^{i\varphi} \quad (-\infty<y<\infty)
\label{A6}
\end{equation}
and $|0 \rangle$ is the ground state of a
specific harmonic oscillator, ${\bf a}\ket{0}=0$. (i.e. {\bf a} is the 
annihilation operator with a definite frequency $\omega_0$;
In this paper, we set $\hbar=m=\omega_0=1$.)  In \rf{A6}  
$\Db(p,q)$ is the displacement operator \rf{d}, which generates
coherent states when applied to $\ket{0}$,  and
\begin{equation}
{\bf S}(\zeta) =\exp \left[ \frac 1{2} \left(\zeta {\bf a^{\dagger 2}-
\zeta^* a^{2}}\right) \right]~, \quad  \left( {\bf a} := \frac
{\Qb +i \Pb} {\sqrt{2}} \right)
\end{equation}
is the {\em squeezing operator}, where the {\em squeeze parameter} $y$
vanishes in the coherent-state limit. 

If $A$ is a density matrix $\rho$, 
then its Q function $Q(\rho:p,q;\zeta)$ can  naturally be
interpreted as a probability distribution. To emphasize this fact,
the Q functions were denoted by $P$ in \cite{md,dab}, instead of $Q$ here.

For simplicity, I shall from now on discuss only
squeezings which are pure boosts, without rotation, i.e. with 
$\varphi \equiv 0$ in \rf{A6},
and use the {\em squeezing parameter} $\l:= e^{2y}$ instead of $y$.

\section{Non-existence of a 1-sided Propagator}

I shall now explain why $G_0$, the Fourier transform of 
the kernel $K$, can not exist. First, we note that the $G_0$ in \rf{ftk}, 
if it exists, would be a special case, {\em for} $\mu=\lambda$ (!),
of the following product of two $G_1$ factors:
\be G(p-p',q-q';\l^{-1}-\lambda^{-1},\l-\mu) :=
G_1(p-p';\l^{-1}-\lambda^{-1}) \; G_1(q-q';\l-\mu)~,  \el{prop}
where
\br G_1(p-p';\l^{-1}-\lambda^{-1}) &:=& 
 \int_{-\infty}^\infty \frac{dx}{2 \pi} e^{i x (p-p')} 
e^{- \frac{x^2}{4}(\l^{-1}-\lambda^{-1})} \nn \\ 
&=& \frac{1}{\sqrt{\pi(\l^{-1}-\lambda^{-1})}} 
e^{-\frac {\; (p-p')^2}{\l^{-1}-\lambda^{-1}}}~,\; \; \qquad  \mbox{
only for} \quad 
\lambda > \l~, \lb{pp} \\ 
G_1(q-q';\l-\mu) &:=&
 \int_{-\infty}^\infty \frac{dk}{2 \pi} e^{i k (q-q')} 
e^{- \frac{k^2}{4}(\l-\mu) } \nn \\
&=& \frac{1}{\sqrt{\pi(\l-\mu)}} 
e^{-\frac {\; (q-q')^2}{\l-\mu}}~, \qquad \qquad \qquad \mbox{only for}
 \quad \mu < \l~.
\erl{pq} 
Note that $G_1(x-x';\tau)$ is the propagator of the one-dimensional 
diffusion equation 
\be \left [ \pd{}{\tau} - \frac{1}{4} \frac{\partial^{2}}{\partial x^{2}}
\right] G_1(x-x';\tau)=0, \for \tau >0~. \ee
The integral \rf{pp} exists only for
$\l < \lambda$, and then it yields the p-propagator,
but it diverges for $\l > \lambda$. In contrast, the corresponding integral
\rf{pq} exists only for $\l > \mu$, but diverges for 
$\l <  \mu$. Thus, {\em their product \rf{prop} cannot exist 
simultaneously, if $\mu=\lambda$}. Consequently,
the desired propagator $G_0$, as a Fourier transform of $K$, cannot exist. 

\section{An algorithm for squeezing Q functions}

To overcome the above difficulty we recall \cite{dab} that the solutions 
of the pseudo-diffusion equation \rf{psd} can be obtained by the 
method of separation of variables: Writing the solution as a product of two
functions, $Q(p,q;\l)=\theta(p,\l)\psi(q,\l)$, where $\theta$
depends only on $p$ and $\l$, and $\psi$ depends only on $q$ and 
$\l$, we get
\br
0 &=& \frac 1{Q} \heartsuit Q = \frac 1{\theta \psi} \left(
\frac{\partial}{\partial \l} -\frac{1}{4}\left[
\frac{\partial ^2}{\partial q^2} -\frac{1}{\l^2}
\frac{\partial ^2}{\partial p^2} \right]\right)\;\theta \,\psi \nonumber  \\
&=&\frac 1{\psi}\left( \frac{\partial }{\partial \l} -\frac{1}{4}
\frac{\partial ^2}{\partial q^2} \right ) \psi(q;\l)- \frac
1{\theta}\left(- \frac{\partial }{\partial \l} - \frac{1}{4 \l^2}
\frac{\partial ^2}{\partial p^2} \right) \theta(p;\l)~.  \lb{psd3}
\er
Since the first term in \rf{psd3} depends only on $q$ and $\l$,
while the second term in (\ref{psd3}) depends only on $p$ and $\l$,
we conclude that each of them must be equal to a function of $\l$ only,
which we denote by $f(\l)$. But I shall consider here  only
the  case $f(\l) \equiv 0$, since it turns out  \cite{dab} that
$f(\l)\ne 0$ yields no new solutions of \rf{psd}. 
For $f(\l) \equiv 0$, Eq. \rf{psd3} 
yields the following two equations:
\br
\left( \frac{\partial }{\partial \l}  - \frac{1}{4}
\frac{\partial ^2}{\partial q^2} \right) \psi(q;\l)&=& 0 \label{inh1} \\
\left(- \frac{\partial }{\partial \l} - \frac{1}{4\l^2}
\frac{\partial ^2}{\partial p^2} \right)\theta(p;\l)= \frac{1}{\l^2} 
\left(\frac{\partial }{\partial \l^{-1}} -
\frac{1}{4}\frac{\partial ^2}{\partial p^2} \right) \theta(p;\l)&=& 0~,
\lb{inh}
\er
where we used $\frac{\partial }{\partial \l} = - \frac 1{\l^2}
\frac{\partial }{\partial \l^{-1}} $ in \rf{inh}.
We see that $\psi$ obeys a 
1-dimensional diffusion equation in $q$, where $\l$ plays the role of
time. Similarly, $\theta$ obeys a diffusion
equation in $p$, but with $ \l^{-1}$ playing the role of time. 

Let us for definiteness discuss the case $\l>1$ (the $\l <1$ case
can be dealt with by similar arguments): For $\l>1$ we can obtain 
the solution $\psi
(q,\l)$ by  using the propagator
\rf{pq}. The action of the propagator for $\l>1$ can also be written 
formally as a differential operator $\Yb_1(\l-1,q)$, acting on the 
unsqueezed function $\psi(q,1)$: 
\br \psi(q,\l) &=& \int dq' \frac{1}{\sqrt{\pi(\l-1)}} 
e^{-\frac {\; (q-q')^2}{\l-1}} \psi(q',1)  \lb{pqc}\\
&=& \exp \left[\frac{(\l-1)}{4} \pd{\,^2}{q^2}\right] \psi(q,1) =:
\Yb_1(\l-1,q) \psi(q,1)~. \er
The propagator in \rf{pqc} will blow up for $\l<1$, and the integral in
\rf{pqc} will not exist, even if $\psi(q,1)$ is a strongly decaying 
function, such as $\psi(q,1)= \exp [- q^2/\alpha]$, for $1-\alpha<\l<1$: 
This is because, the coefficient of $q'^2$ in the exponent in the 
integrand of \rf{pqc} would be positive: 
$\frac{1}{1-\l}-\frac{1}{\alpha}=\frac{\alpha-1+\l}{\alpha(1-\l)}>0$.
In contrast, if we {\em first evaluate the integral \rf{pqc} using} $\l>1$, 
{\em then we can continue the resulting function} $\psi(q,\l)$ {\em to} 
$1-\alpha < \l <1 $, as we shall illustrate below.

This suggests the following algorithm for calculating the
squeezed $p$ factor: First, calculate the  solution $\theta(p,\l)$
of the diffusion equation \rf{inh} by using the p-propagator for $1/\l >1$. 
Then continue the resulting solution to $1/\l<1$. 
One could also calculate the solution
$\theta(p,\l)$  directly for $1/\l<1$, by using 
the power series expansion of the differential operator 
$\Yb_1(\l^{-1}-1,p)$:
\be \theta (p,\l^{-1})= \Yb_1(\l^{-1}-1,p)\, \theta(p,1)
=\exp\left[(\l^{-1}-1) \pd{\, ^2}{p^2}\right] \theta(p,1)~. 
\ee
As an example, let us squeeze the Q function of thermal light \cite{dab},
whose density matrix is given by:
\be  \rho_{th} = \frac 1{ \overline{n}+1} \sum_{n=0}^\infty 
\ket{n} \left(\frac {\overline{n}}{ \overline{n}+1}\right)^n \bra{n} ~,
\where   \overline{n} \equiv \frac 1{e^\beta-1}~, \el{therm}
where  $\beta=\hbar \omega /k_B T$ and  $\overline{n}$ is the mean number
of photons. The (coherent) Q function of \rf{therm} is factorizable:
\be 
Q(\rho_{th}:p,q;1)
= \frac{1}{\overline{n}+1}\exp \left[ -\frac{p^2+q^2} {2(\overline{n}
+1)}\right]=\psi(q,1)\, \theta(p,1)~,  \label{th}
\ee
where each factor turns out to be proportional to the diffusion propagator: 
\be
\psi(q,1)=\theta(q,1)= \frac 1{\sqrt{\overline{n}+1}}
\exp \left[ -\frac {q^2}{2(\overline{n}+1)}\right]
=\sqrt{2\pi} \, G_1(q,2\overline{n}+2)~.  \el{thf}
Since the unsqueezed factors are the same, 
$\psi(q,1)=\theta(q,1)$, we only need to evaluate one integral,
by using the propagator for $\l >1$:
\br
\psi(q,\l) &=&
 \int dq' \frac 1{\sqrt{\pi(\l-1)}} 
e^{-\frac {(q-q')^2}{\l-1}} \psi(q',1)=
\sqrt{2 \pi} \int dq'  G_1(q-q',\l-1)\, G_1(q',2\overline{n}+2)\nn \\
&=& \sqrt{2\pi}\, G_1(q,\l+2\overline{n}+1)=
 \sqrt{\frac 2{\l+ 2\overline{n}+1}} \exp \left[-\frac{q^2}
{\l+2\overline{n}+1} \right]~.
\erl{pqf}
We see that $\psi(q;\l)$ is well defined for $ -(2\overline{n}+1)<\l <1$.
Since  $\theta(p,1)=\psi(p,1)$, we get  $\theta(p,\l)=\psi(p,\l^{-1})$,
so that the squeezed Q function becomes 
$ Q(\rho_{th}:p,q;\l)=\psi(p,\l^{-1}) \psi(q,\l)$.
Thus, the squeezing of $Q(\rho_{th}:p,q;1)$ required only a single 
integration, which in this specific case was carried out by noting that
the successive action of two propagators is equivalent to the
action of one propagator, whose `time variable' is the sum of the 
individual times: $(\l-1)+(2\overline{n}+2)=\l+2\overline{n}+1$.
 
\section{Forward-Backward Squeezing Propagators}

The products in \rf{prop} will exist and be solutions of the 
pseudo-diffusion equation,
if we relax the condition $\mu=\lambda$  and demand instead
only $\mu <\l < \lambda$. Actually, these products yield a different 
solution for each 4-tupel $(p',q',\lambda,\mu)$. 
Since the heart operator $\hrt$ is linear, any superposition
of these solutions is also a solution. In particular, 
if we fix the squeezing 
parameters $\mu$ and $\lambda$ and integrate only over $p'$ and $q'$, we get
solutions of the form
\be f(p,q;\l^{-1},\l)=\intt dp' dq'\;  
G(p-p',q-q'; \lambda^{-1}-\l^{-1},\l-\mu) \; f(p',q';\lambda^{-1},\mu), 
\for  \mu < \l < \lambda~,\lb{liv}\ee
for any given function $f(p,q;\lambda^{-1},\mu)$, provided that the 
integrals \rf{liv} exist. We see that $G$ in \rf{liv} is acting as 
a propagator, which provides a solution of the 
pseudo-diffusion equation \rf{psd}
for the squeezing parameter $\l$ in the range $\mu <\l<\lambda$.
I shall therefore call these $G$ functions 
{\em two-sided or forward-backward squeezing propagators} of the 
pseudo-diffusion equation 
\rf{psd}, since the two squeezing parameters, $\mu$ and $\lambda$,
lie on opposite sides of $\l$. These $G$ solutions have the 
proper limit, which one expects from a propagator, if $\l$ is 
approached from opposite directions:
\be 
\lim_{\mu \rightarrow \l-\epsilon \; , \; \lambda \rightarrow \l+\epsilon}
 G(p-p', q-q';\lambda^{-1}-\l^{-1},\l-\mu)=\delta(p-p')\; \delta (q-q')~.
 \lb{lim} \ee
An extreme case of the  squeezing propagators \rf{prop} is obtained by 
choosing $\mu=0$ and $\lambda=\infty$. These squeezing parameters 
correspond to the values $-\infty$ and $+\infty$ of the 
$y=\frac 1{2} \ln \l$  variable, respectively: 
\be G(p-p', q-q';\l^{-1},\l) =
\frac1{\pi} \exp \left[- \l(p-p')^2
- \l^{-1}(q-q')^2 \right]~, \for 0 < \l <  \infty~. 
\el{prope}
For the choice $\mu=0$ and $\lambda=\infty$ in \rf{liv}, $\l$ can now
take any positive value. Moreover, the square-root 
factors in the two propagators cancel out. 
Thus, with the propagator \rf{prope}, the relation \rf{liv} becomes 
simply
\be f(p,q;\l^{-1},\l)=\int \! \! \! \int  
\frac{dp'dq'}{\pi} \exp [-\l
(p-p')^2-\l^{-1} (q-q')^2 ]\,f(p',q';0,0), \for  \l> 0~. \lb{liv1}\ee

\section{New Interpretation of the Wigner Function}

The Wigner representation of the projection operator $\Pib(p,q;\l)$
is given by
\br
\Pib(p,q;\l) &=& \intt  \frac{dp'dq'}{2\pi} \Tr [\Pib(p,q;\l) \W(p',q')]
\; \W(p',q') \nn \\
&=& 2  \intt  \frac{dp'dq'}{\pi} \exp [-\l(p-p')^2-\l^{-1} (q-q')^2 ]\; 
\W(p',q')~.
\erl{p1}
where we used 
\br
\Tr [\Pib(p,q;\l) \W(p',q')]\!\!\!&=&\! \!\bra{p,q;\l} \W(p',q')\ket{p,q;\l}
=\bra{p,q;\l}D(p',q')\W(0,0)D^\dagger(p',q')\ket{p,q;\l}\qquad \nn \\
 &=& \!\!\!\!
 \sp{p-p',q-q';\l|p'-p,q'-q;\l}
= \exp [-\l(p-p')^2-\l^{-1} (q-q')^2 ]~. \erl{pw}

By multiplying \rf{p1}
by an operator $A$ and taking the trace, we get a similar
relation between the Q function and its Wigner counterpart \cite{kim}:
\be
Q(A:p,q;\l) = \int \! \! \! \int 
\frac{dp'dq'}{\pi} \exp [-\l
(p-p')^2-\l^{-1} (q-q^{\prime
})^2 ]\, W(A:p',q'),  \lb{rel} \ee
Comparing this relation with \rf{liv1}, 
we realize that the Q function is a propagated Wigner function via
by the above special 2-sided squeezing propagator \rf{prope}.
This led me to define
an interpolating function $Q(A:p,q;\l_p^{-1},\l_q)$ in Eq. \rf{qw} below, 
which for $ \l_q=\l_p=\l$ 
yields the usual squeezed Q function
$Q(A:p,q;\l)$, whereas for $\l_p=\infty$ and $\l_q=0$,
it yields the corresponding Wigner function. Hence, we can interpret {\em 
the Wigner function as a generalized Q function, which is 
squeezed to `infinite future' $y_p=+\infty$ ($\l_p = e^{2y_p}=\infty$) in 
the $p$ variable, but anti-squeezed to `infinite past' $y_q=-\infty$
($\l_q = e^{2y_q}=0$) in the $q$ variable.}

\section{Generalized Q functions}

The above interpretation suggests the following definition of
{\em generalized Q functions} which
depend on two different squeezing parameters, $\l_q$ and $\l_p$:
Since the Wigner functions correspond to infinite past and infinite
future, we can use the 2-sided propagators to bring them into
finite past $\l_q$ and finite future $\l_p$ (or ``two different points of
the present"), as follows 
\br Q(A:p,q;\frac 1{\l_p},\l_q) &:= & \intt
\frac{dp'dq'}{\pi} G_1(p-p';\l_p^{-1})\, G_1(q-q';\l_q)\; 
Q(A:p',q';0,0) \\
 &= &  
\sqrt{\frac{\l_p}{\l_q}} \intt
\frac{dp'dq'}{\pi} \exp [-\l_p (p-p')^2-\l_q^{-1} (q-q')^2 ]\, 
W(A:p',q')~,  \erl{qw}
where  $W(A:p,q) = Q(A:p,q;0,0)$ was used.
Note that, for now, we assume $\l_q, \l_p> 0$ in \rf{qw}, but $\l_p$ 
may be larger or smaller than $\l_q$. Later, we shall extend the range
of  $\l_q$ and $\l_p$. 

Instead of studying Q {\em functions}, it is more useful 
to  define and study generalized \Qb {\em operators}, which yield  Q
functions by taking the trace $\Tr (A \Qb)$.
Similar to \rf{qw}, we generalize the operator relation \rf{p1},
as follows:
\br
\Qb(p,q;\l^{-1}_p,\l_q) &:=& 
2 \ \exp \left[\frac{1}{4}\left(\frac{1}{\l_p} \frac{\partial^{2}}
{\partial p^2}+\l_q \pd{^{2}}{q^2}
 \right) \right] \; \W(p,q)  \lb{gw2} \\
&=& 
2\ \sqrt{\frac{\l_p}{\l_q}}\intt 
\frac{dp'dq'}{\pi} \exp [-\l_p (p-p')^2-\l_q^{-1} (q-q')^2 ]\;
 \W(p',q')  \lb{gp1}\\
&=&  \intt 
\frac{dk dx}{2 \pi} 
\exp \left[-\frac{1}{4}(\frac{1}{\l_p} x^2+\l_q k^2)+ i(px-qk)\right]
\Db(k,x)~, \erl{gft}
where \rf{gft} follows from \rf{gp1} by substituting 
the Fourier transform of the Wigner operator
\cite{dahl}:
\be
\W(p, q) =\frac 1{2} \intt \frac{dk dx}{2\pi} e^{i[px-qk]}\D(k,x)~.
\el{ivw3}
So far we assumed that $\l_p$ and $\l_q$ are positive. However,  
we can formally extend the  definition \rf{gw2} to real (and even 
complex) values of $\l_p$ and $\l_q$.

From \rf{gw2} or \rf{gft} we see immediately that the \Qb operators 
(and hence also the generalized Q functions via the trace) satisfy the 
following partial differential equation
\be \left [ \pd{}{\mu} - \frac{1}{4} \left(\pd{\l_q}{\mu}
 \frac{\partial^{2}}{\partial q^{2}}
- \frac{1}{\l_p^{2}} \pd{\l_p}{\mu} \frac{\partial^{2}}{\partial p^2}
 \right)\right] \Qb(p,q; \frac 1{\l_p(\mu)}, \l_q (\mu))=0~,   \lb{gpsd2}
\ee
{\em if $\l_q$ and $\l_p$ are varied along a curve} $(\l_q(\mu),\l_p(\mu))$
{\em in the $(\l_q,\l_p)$ plane}. We see that \rf{gpsd2} yields
a generalized pseudo-diffusion (diffusion) equation, if $\pd{\l_q}{\mu}$ and
$\pd{\l_p}{\mu}$ have the same (opposite) signs. In particular, 
on the subdomain of $\R^2$, defined by
\be  
 \l_p= \l \t^{-1}~,\qquad \l_q= \l \t ~,  \where  \l>0 \, , \, \t >-1~.
\el{newv}
the operator $ \Qb(p,q;\t \l^{-1},\t \l)$ yields a generalization of 
the Cahill-Glauber
interpolation operator \cite{cahill}, which corresponds to zero squeezing 
($\l\equiv 1$):
\be  
\Qb(p,q;\t\l^{-1},\t \l) =
\intt \frac{dk dx}{2 \pi} 
\exp \left[-\frac{\t}{4} ( \l^{-1} x^2+\l k^2) + i(px-qk)\right]
\Db(k,x)~.  \el{gtl} 
We see that the parameter $s$ of \cite{cahill} corresponds to our 
$-\t$ here.
I shall call $\t$ the {\em thermal variable}, since $\Qb(0,0;\t,\t )$
is equal to the thermal density operator $\rho_{th}$ in \rf{therm},
if we make the identification $\t=\coth (\beta/2)$. Hence, 
$\Qb(0,0;\t,\t )$ is
equal to the thermalized parity operator of \cite{vb}
and to the generalized parity operator of \cite{ben}, which was
introduced in order to obtain the parity operator $\W(0,0)$
as a limit of trace-class operators.

From equation \rf{gpsd2} we see that $ \Qb(p,q;\t \l^{-1},\t \l)$
obeys the following two interesting partial differential equations: 
\br 
\heartsuit(p,q;\t,\l)\   \Qb(p,q;\t \l^{-1},\t \l) :=  
\left [ \pd{}{\l} - \frac{\t}{4}
 \left( \frac{\partial^{2}}{\partial q^{2}}
- \frac{1}{\l^2} \frac{\partial^{2}}{\partial p^2} \right)\right]
  \Qb(p,q;\t \l^{-1},\t \l)
& =& 0~,   \lb{gpsd}\\
 \left [ \pd{}{\t} - \frac{1}{4}
 \left(\l \frac{\partial^{2}}{\partial q^{2}}
+ \frac{1}{\l} \frac{\partial^{2}}{\partial p^2} \right)\right]
 \Qb(p,q;\t \l^{-1},\t \l) &=& 0~,   \lb{tpde} 
\er
if $\l$ is varied along {\em `isothermal curves'}
($ \l_q/\l_p=\t^2={\mbox{const.}}$), or  $\t$ varied along the 
{\em `isosqueeze 
curves'} ($ \l_q\l_p= \l^2={\mbox{const.}}$), respectively.

Eq.\rf{gpsd} is a {\em generalized pseudo-diffusion equation},
since $\heartsuit(p,q;\t,\l)$ now depends also on 
$\t$ and reduces to the original heart operator 
in \rf{psd} for $\t=1$. For  $\t=0$ equation \rf{gpsd} tells us that 
{\em the} \Qb {\em operator is not affected by squeezing}: 
$ \Qb(p,q;0,0)= 2 \W(p,q)$ for all $\l$.

Eq.\rf{tpde} is a diffusion equation in 
2 dimensions \cite{widder}, with
different diffusion constants in the $p$ and $q$ directions, if $\l\ne 1$.

\section{Summary}

I reviewed the 3-steps Fourier-transform procedure for calculating squeezed 
Q functions $Q(A:p,q;\l)$ from given unsqueezed or less squeezed ones
$Q(A: p,q;\mu)$, where $\mu < \l$ \cite{dab}:
First calculate $\Qt(A:k,x;\mu)$, the Fourier transform of $Q(A:p,q;\mu)$, 
then  multiply $\Qt$ by the special kernel $K$ and 
finally take the inverse Fourier transformation 
of the product $K\Qt$ and you get $Q(A:p,q;\l)$. 
Instead, I gave an algorithm for squeezing Q functions directly, by 
using one-dimensional diffusion operators.

I also explained why one-sided propagators cannot be defined for
 the pseudo-diffusion equation \rf{psd} and showed that 
two-sided squeezing propagators 
\rf{prop} are more appropriate for this equation. 

I  noted that the Q functions are related to their  Wigner
counterparts by the extreme two-sided propagator \rf{prope}, and concluded 
that the Wigner functions can be looked upon as generalized 
Q functions $Q(A:p,q;\l_p^{-1},\l_q)$, which are squeezed forwards 
($\l_p=\infty$) in $p$ variable and backwards ($\l_q=0$) in the $q$ 
variable.
 
These $\Qb$ operators were defined formally
in \rf{gw2} in terms of Wigner operators. 
On a subdomain of the parameters
($\l_p,\l_q$), which is defined in terms of $\t$ and $\l$ in \rf{newv}, 
these \Qb operators yield the `thermalized and squeezed 
Wigner operator' $\Qb(p,q;\t\l^{-1},\t \l)$, which  yields
the interpolating operator of Cahill and Glauber \cite{cahill}
for zero squeezing ($\l \equiv 1$). However,
we emphasize that even $\Qb(p,q;\t\l^{-1},\t \l)$ for $\l \ge 0$ is 
less general than our generalized \Qb operator 
$\Qb(p,q;\l_p^{-1},\l_q)$: For example, 
$\Qb(p,q;0,\l_q)$ does not correspond to a thermalized and squeezed 
Wigner operator.

I also showed that $\Qb(p,q;\t\l^{-1},\t \l)$ obeys 
the generalized pseudo-diffusion equation \rf{gpsd} for constant $\t$,
and the 2-dimensional diffusion equation \rf{tpde} for constant $\l$.
This diffusion equation  \rf{tpde} explains intuitively 
why each of the P, the Wigner and the Q distributions, become smoother 
than its predecessor, as $\t$ is {\em increased} from  $\t=-1$ to $\t=0$ to 
$\t=1$, for any $\l={\mbox{const.}}$ 

This smoothing process is expected to continue as $\t$ is increased from 
$1$ to $\infty$. Indeed, it can be shown that the generalized Q functions
$Q(\rho:p,q;\t \l^{-1},\t \l)=\Tr\ [\rho \Qb ( p,q;\t \l^{-1}, \t \l )]$
of density matrices $\rho$ are nonnegative for $\t \ge 1$, and thus they 
yield {\em probability distributions} and not merely quasi-probability
distributions. Consequently, their Wehrl entropy \cite{wehrl}
\be S(\t,\l):=-\intt \frac {dp dq}{2\pi} Q(\rho:p,q;\t \l^{-1},\t \l)
\ln Q(\rho:p,q;\t \l^{-1},\t \l)~,  \for \t \ge 0~, \ee
is well defined and should increase if we vary both $\t$ and $\l$ 
simultaneously in such a way that both $\l_q$ and $\l_p$ increase. 
Hence,  further clarification of the physical meaning of 
the {\em observable} $\Qb(p,q;\t \l^{-1}, \t \l)$ would be useful.

\end{document}